**Abstract**

The goal of this research is to uncover the channels through which research and development (R&D) impacts economic growth in developing countries. The study employed nine variables from three broader categories in the World Economic Forum database, each covering 32 countries from the lower-middle-income group for the year 2019. The theoretical framework is based on the R&D ecosystem, which includes components such as Institutions, Human capital, Capital market, R&D, and Innovation. Each of these components can contribute to the economic development of the country. Using Structural Equation Modelling (SEM), we build a path diagram to visualize and confirm a potential relationship between the components. R&D features had a positive impact on innovation (regression weight estimate: +0.34, p = 0.001), as did capital market institutions (regression weight estimate: +0.12, p = 0.007), but neither had a significant impact on growth. According to the Schumpeterian institutional interpretation, R&D and innovation efforts may not lead to sustained growth in middle-income countries. We find no significant connection between innovation performance and economic growth. This suggests that while R&D and capital markets may contribute to innovation through entrepreneurship, this contribution is not impactful enough to drive economic growth in developing countries. Our findings provide further evidence of the middle-income trap.

# 1. Introduction

New Growth Theory (NGT) commonly refers to the series of endogenous growth theories proposed by various researchers. According to Capolupo (2009), change in growth can be accounted for by endogenous technical factors such as learning by doing (Romer, 1986); spillovers by human capital (Lucas, 1988); threshold externalities (Azariadis & Drazen, 1990); production externalities (Barro, 1990) and improving quality through invention (Grossman & Helpman, 1991). NGT states that long-term growth can only be attained by knowledge generation (Ge & Liu, 2022). If we acknowledge that there are beneficial externalities that "compensate" for the declining marginal productivity of capital assets, we may also observe sustainable growth. These externalities are the result of projects like public infrastructure building, knowledge distribution and R&D. To put it briefly, because the returns from the accumulation elements are constant, growth is a self-maintaining phenomenon that occurs at a constant rate (Diebolt & Perrin, 2019).

For developing economies, NGT—and especially the growth debate it sparked—has provided some insightful lessons. First of all, it is now evident that the per capita income levels of developing countries can either converge with or deviate from those of developed economies.

The actions and conditions of the developing economies themselves will determine the result. Therefore, NGT contributes to dispersing the myth that Neo-Classical Growth Theory is irrelevant to policy in terms of the long-term growth rate (Lee, 2020). The empirical research spurred by NGT has revealed significant productivity disparities between economies, prompting developing economies to focus on productivity rather than just investment volume (Rizwanul, 2004). Moreover, NGT focuses on the function of institutions in economic development. With the recent emergence of data sets measuring different dimensions of institutional quality, it is now possible to quantitatively analyze how institutions have contributed to the development of emerging economies. NGT makes the critical link between the development of human capital and the spread of institutional and technological innovations more apparent. It also highlights the challenge facing developing economies in building up their human capital (Prasetyo, 2020).

The purpose of this study is to uncover the channels through which R&D impacts the economy of a developing country. Numerous studies have found that this impact is influenced by other factors in the economy, including institutions (Doloreux & Turkina, 2023; Chen & Song, 2024). Institutions play a key role in the stability of an economy as they regulate different sectors of an economy. Our primary aim is to investigate whether R&D performance has an impact on growth among developing countries. It is suggested by Aghion & Bircan (2017) that the institutional framework could slow growth down at a certain level of development. Due to these institutional structures, R&D and innovation may not lead to economic growth in middle-income countries. We test the relationship between institutions, R&D, innovation, and growth in a lower middle-income group of countries using structural equation modelling. Our research contributes to the literature discussing the role and the limitations of R&D and innovation in economic growth.

The remainder of this paper is structured as follows: Sections 2 and 3 present the theoretical framework and supporting literature. Section 4 describes the data and statistical methods employed. Section 5 presents the empirical results, followed by a discussion of the findings and their implications in Section 6. Finally, Section 7 provides concluding remarks.

## 2. Literature review

R&D is described as a methodical and creative activity aimed at growing the stock of knowledge (including an understanding of society, culture, and people) and creating new applications for existing information (OECD, 2015). Our research tests the interaction between R&D and innovation, growth, institutional factors related to human capital accumulation, and capital markets. Therefore, a separate review subchapter is devoted to each of these interactions.

### 2.1. R&D expenditures and economic growth

Surpassing the technical growth method found in the models of Solow (1956) and Mukerji and Johansen (1963), the Romer (1989) growth model demonstrates how technology can promote self-maintained growth. He argues that technological knowledge is universally accessible and benefits from increased research investment, which accelerates technological production. This industry experiences rising returns as past discoveries enhance researcher productivity. Research generates positive externalities by creating new capital assets and boosting future researcher productivity, benefits not reflected in market prices. These externalities are linked to information distribution (Diebolt & Perrin, 2016).

Coe et al. (2009) assert that R&D investment is critical for economic growth. Aghion et al. (2011), using US data, discovered in their empirical analysis that R&D spending as a proportion of GDP has an effect on economic growth. Lichtenberg (2001) examined the impact of public and private sector R&D spending on economic development, using data from 74 countries between 1964 and 1989. He concluded that while R&D investment benefits growth, private sector R&D spending is more efficient and effective than public sector R&D spending. Coe et al. (1997) estimate the spillover effect of wealthy countries' R&D expenditure on developing countries. They assert that R&D investments in wealthy nations not only benefit their own economies but also those of developing countries. They further argued that high-tech commodities, as well as capital goods exported from developed countries to developing ones, boost the efficiency of developing countries. They found that increasing R&D expenditure in developed countries has a large positive effect on output in developing countries. However, Ke (2024), investigating Malaysia, Thailand, Indonesia, and the Philippines, concludes that these countries are caught in a "middle technology trap" with limited technology spillovers. Lichtenberg and Pottelsberghe de la Potterie (1998) studied the effect of R&D spending on the rise of total factor productivity (TFP) in developing nations. Their research casts doubt on the findings of Coe et al. (1997) and points out that additional resources should be directed to research and development in order to attain long-run economic growth performance.

## 2.2. Human capital

Lucas (1988) examines the individual choices made in the pursuit of knowledge and how those choices affect both individual productivity and overall economic progress. He argues that human capital, acting as a catalyst for growth, serves as both an alternative and a supplement to technological advancement. In Azariadis and Drazen's (1990) endogenous growth model with interconnected generations, human capital is the growth engine. This is because it accumulates and shows growing social returns to scale. In the model, all individuals are identical, and their lives may be split into two phases. The first is a time of training and employment. The following phase is a time of working alone (the time spent in education in their early years is translated into later high-quality labor). Growing social returns are also generated when they unintentionally leave a portion of their human capital to their successors when they pass away (Altinok & Diebolt, 2024).

Empirical studies by Barro (1992) and Barro et al. (1995) have demonstrated that the economies that have been most successful in catching up to industrialized economies are those which possess a higher degree of human capital in relation to their starting income. A relationship between human capital and adoption of new technologies was suggested by Nelson and Phelps (1966) and, more recently, by Färnstrand Damsgaard and Krusell (2010). Benhabib and Spiegel (2005) empirically demonstrated that human capital has two significant functions for TFP growth. If human capital generates production via two sources – innovation and knowledge adaptation – then measuring R&D spillovers presents a conceptual challenge. The concept of R&D spillover in Griliches (1998) and Romer (1986) is limited to external data used in enterprises' and organizations' formal R&D operations. The objective of this R&D is to create new products and inputs and to improve manufacturing processes. Endogenous growth theory is expressly limited to formal research and development as the means of TFP as well as growth initiative. Learning extends beyond the walls of R&D institutions and contributes to individual competencies that competitors can utilize. This type of learning contributes to the human capital pool and, consequently, improves factor quality. However, it is not captured by the TFP residual.

Krueger (1968) demonstrated through a cross-country analysis that the most significant factor in explaining the variations in per-capita income between the United States and emerging nations was disparities in human capital. Complex growth accounting investigations, like those of Jorgenson et al. (2016), have demonstrated a significant contribution of human capital accumulation to growth even within the rigorously neoclassical framework (Mincer, 2022). Devassia et al. (2024) carried out research on a sample of countries that included Northern Macedonia and found that while general education has a positive impact on growth, an increase in the tertiary enrolment ratio actually hurts growth.

## 2.3. Capital market

Capital markets are crucial because they provide capital for new enterprises. Entrepreneurship is key for innovation, driving the conversion of knowledge, and ideas into products and GDP. Will (2019) defines the capital market as the system that directs savings into long-term investment. A robust financial sector capable of pooling domestic resources and attracting foreign capital is crucial for economic growth. Weak financial institutions hinder investment, limit capital accumulation necessary for global competitiveness, and ultimately reduce potential growth. Underdeveloped capital markets also deter foreign investment due to high costs and illiquidity. Empirical research by Shaw (1973) and Degong et al. (2021) provides evidence of a positive correlation between growth and liberal financial policy.

## 2.4. Institutions

Institutions play an important role in affecting a country's level of economic growth. Cross-country studies show that metrics such as the degree to which property rights are protected, the law system, and civil liberties are strongly related to economic performance. The literature has described why institutions are so important for economic advancement and has provided facts to support its claims (e.g.., Knack & Keefer, 1995; Son, 2016; Cvetanović et al., 2019).

North (1991) defines institutions as "the rules of the game in a society, the consciously constructed limits that influence human interaction." They shape human transaction incentives, whether they be political, societal, or economic. Acemoglu and Robinson (2005) argue that while "strong" institutions, such as property rights, can drive structural change, their effectiveness depends on proper implementation within the existing economic structure. They caution that even when implemented, these institutions do not guarantee long-term economic growth. Studies by Acemoglu et al. (2001) demonstrate that institutional attributes are more effective in the long run than in the short run. There is evidence that institutional factors affect TFP and that nations with superior institutions have higher productivity (Méon & Sekkat, 2004).

The institutional structure may weaken innovation in developing countries, leading to a middle-income trap (Aghion & Bircan, 2017). Lock-ins of political and economic power structures of the old elites can slow down the rate of growth and may prevent R&D and innovation efforts from disrupting the markets (Mickiewicz, 2023). The old structures may also prevent the government from investing in institutional capacity needed for human capital investments (Doner & Schneider, 2016). These limitations could also explain why it is so difficult to set up a functional theory of economic growth in middle-income economies (Kharas & Gill, 2020).

## 2.5. Innovation

Innovation, the mechanism by which new information is applied to economic processes, catalyzes long-term economic progress (Smith et al., 2012). Experimental studies have demonstrated that business innovation—particularly functional enhancements in processes, systems, services, and products—is essential for both firm competitiveness and overall economic growth (Smith & Estibals, 2011). Innovation requires considerable expense: it takes financial investment and resource commitment. Investment in tangible and intangible innovation assets is necessary to foster productivity growth, benefitting both GDP growth and overall prosperity (Smith et al., 2012). Durham (2004) looked into the connections between national R&D initiatives and innovation, as well as the relationship between innovation and per capita income. The investigation examined R&D and patent data for 20 OECD and non-OECD nations from 1987 to 1997, using a variety of panel data approaches. The findings indicated a positive correlation between innovation (patent stock) and per capita GDP in both OECD and non-OECD nations. However, only larger OECD nations were able to boost their level of innovation through R&D spending.

A coherent innovation strategy must adapt to changing global environment. In this differentiated environment, economic gains can be realized through the development of new technologies and organizational forms. These gains also come from adapting and deploying these innovations to produce new goods and services across the spectrum of existing economic and social activities (Brunswicker & Schecter, 2019).

## 3. Theoretical framework and hypothesis development

The ecosystem literature encompasses a wide range of stakeholders participating in the value chain. This wide range of stakeholders is reflected in the literature's scope and varied levels of analysis. Initially, there are the elementary value chain players, for example, the central firm as well as its customers and suppliers (Moore, 1993; Iansiti & Levien, 2004), extending to state research institutions and universities (Van der Borgh et al., 2012; Clarysse et al., 2014), public organizations (Thomas & Autio, 2014; Li & Garnsey, 2014), and other supporting organizations.

This study, focusing on R&D, assumes that it has a rather indirect impact on economic growth. Various studies have identified multiple factors affecting the economic growth of a country. These studies also indicate that R&D is influenced by institutions, which play a major role in the regulation, implementation, and execution of policies related to all sectors of the economy. Institutions affect the R&D sector, capital markets, and human capital, which in turn affect the innovation capability of a country. Consequently, as the innovation capability is affected, so is the economic growth of the country. On the basis of these theoretical findings, in this study we call this impact cycle the "R&D ecosystem". The term R&D ecosystem was developed based on a synthesis of research on the different types of ecosystems. The three types of institutions in this ecosystem that will collectively contribute to the economic development of the country are:

1. **R&D Institutions (RDI):** A proposed term to refer to the organizations (public or private) that focus on scientific research, technological development, and innovation in order to produce new or improve the quality of existing products and services they offer. It includes organizations such as educational institutions, public or private companies and non-profit organizations. The government plays a crucial role in facilitating research and development within a country by allocating funds appropriately to maximize resource utilization.

2. **Financial Institutions (FI):** Refers to the organizations that are involved in different financial transactions including savings, loans, deposits, investments, etc. in a country. It includes capital markets such as stock markets, banks (commercial and retail), brokerage companies, insurance companies/agencies, credit loan associations/firms, etc.
3. **Labor Institutions (LI):** Refers to the organizations that determine the quality and education level of the labor force, as well as policies regarding employment and wage-related factors.

Figure 1 shows the R&D ecosystem model our research uses.

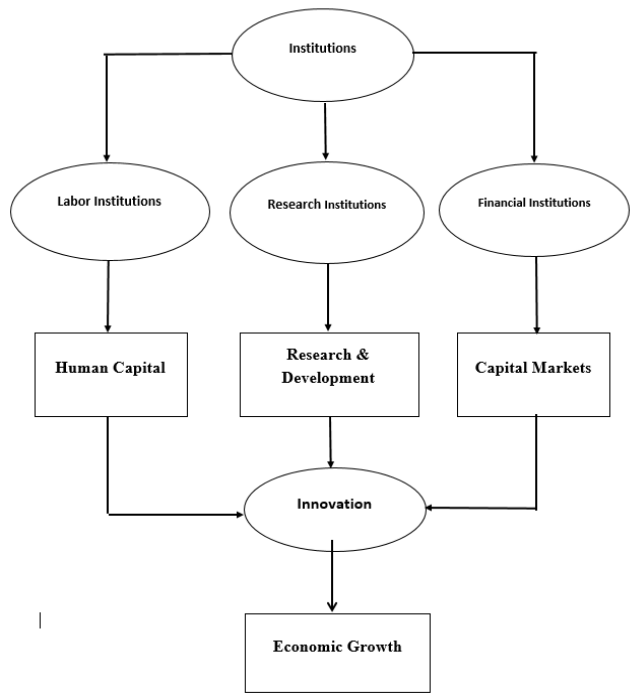

*Figure 1: The R&D ecosystem (theoretical framework)*
*(Source: own work)*

Based on the theoretical framework the following hypotheses can be deduced:

*H1: Institutions related to R&D performance have a significant impact on innovation*

*H2: Institutions related to capital market operations have a significant impact on innovation*

*H3: Institutions related to human capital formation have a significant impact on innovation*

*H4: Innovation has a significant impact on economic growth*

## 4. Materials and methods

The data for this study are from 2019, gathered from the World Economic Forum (WEF) databases and reports available on the organization's official website. The research focuses on the economic and strategic shortcomings of the countries before the onset of the Covid-19 pandemic in early 2020. Therefore, only data from 2019 were used in this study.

Data for 141 countries are included in the WEF database. We selected the lower-middle-income group of countries (as classified by the World Bank Atlas method) because our research aims to uncover the impact of R&D in developing countries. Secondly, almost all the countries in this group are exposed to similar (if not identical) economic, political, and legal environments. The group is comprised of 32 countries (see Table 2).

### 4.1. Variables

This research develops six main constructs: Economic growth, R&D, Institutions, Human capital, Capital market and Innovation. Economic growth, the dependent variable, is measured as the 10-year average annual GDP growth percentage (valued at Purchasing Power Parity – PPP, in billions of USD) from 2009 to 2018. Of the six constructs, three (Aggregated constructs: Economic growth, Institutions and Innovation) are adopted from the WEF database. The remaining three (Component variables: Research & Development - RDI, Capital market - FI, and human Capital - LI) are computed using factor analysis from components provided by the WEF.

**Aggregated constructs**

WEF uses an aggregation method to calculate indicators for its Global Competitiveness Index (GCI). Scores are calculated from the most disaggregated (indicator) level to the most aggregated (highest) level. Each aggregated level is made up of several components. The overall score for each aggregated level is the arithmetic mean (average) of its component indicators. Before aggregation, the raw values of individual indicators are transformed into progress scores ranging from 0 to 100, with 100 being the ideal state (Table 1).

Table 1: The dependent variable (GDP growth) and the aggregated variables of the research
*(Source: Schwab, 2019)*

| Major indicator | Abbreviation (used in Figure 2) | Definition | Scale | Periodicity |
|---|---|---|---|---|
| **GDP growth** | GDP | 10-year average annual GDP Growth (%, real terms) | Weighted Average | Annual, 2009-18 |
| **Institutions** | Ins | Regulate the setting in which the economy and society organize themselves by formal and informal checks and balances | Score: 0-100 (0=worst, 100= Frontier) | Annual, 2019 |
| **Innovation** | Inn | The conversion of new ideas into product and services | Score: 0-100 | Annual, 2019 |

**Component variables**

Our R&D, Capital market, and Human capital constructs are each built from three different indicators. These variables are abbreviated in our model calculations as RDI, FI, and LI. All components are adopted from the World Economic Forum (Schwab, 2019). Table 2 presents descriptive statistics for the variables across the 32 selected countries.

- *Research & Development (RDI):* Scientific publications (ScP); Research institution prominence (RIP); R&D expenditure (RDE)
- *Capital Market (FI):* Domestic credit to private sector (DCPS); Financing of SMEs (FSME); Venture Capital Availability (VCA)
- *Human Capital (LI):* Mean years of schooling (MYS); Quality of vocational training (QVT); Digital skills among active population (DSAP)

Table 2: Descriptive statistics of the variables for the 32 selected lower-middle-income countries

*(Source: Own calculations based on Schwab, 2019, World Bank, 2020)*

| Variable | N | Minimum | Maximum | Mean | Mode | Median | Std. Deviation |
|---|---|---|---|---|---|---|---|
| **GDP - Average annual GDP growth 2011-20** | 32 | .10 | 6.90 | 4.1188 | 2.20 | 3.9 | 1.65499 |
| **Ins - Institutions** | 32 | 36.40 | 60.00 | 47.4938 | 41.90 | 47.85 | 6.13588 |
| **Inn - Innovation** | 32 | 18.80 | 50.90 | 31.7375 | 30.70 | 30.85 | 5.89323 |
| **ScP - Scientific publications** | 32 | 47.10 | 92.70 | 69.9250 | 47.10 | 69.95 | 9.40391 |
| **RIP - Research institutions prominence** | 32 | 0.00 | 98.40 | 5.9625 | 0.00 | 0.55 | 17.44269 |
| **RDE - R&D expenditure** | 32 | 0.50 | 84.00 | 12.8844 | 11.10 | 10.85 | 14.78107 |
| **DCPS - Domestic credit to private sector** | 32 | 2.90 | 100.00 | 42.3656 | 15.8 | 41.40 | 23.61560 |
| **FSME - Financing of SMEs** | 32 | 21.30 | 61.00 | 41.9406 | 21.30 | 43.50 | 9.47958 |
| **VCA - Venture Capital Availability** | 32 | 11.80 | 52.70 | 29.0750 | 19.60 | 27.85 | 9.77776 |
| **MYS - Mean Years of Schooling** | 32 | 18.90 | 77.20 | 47.2313 | 46.00 | 46.00 | 14.29467 |
| **QVT - Quality of vocational training** | 32 | 25.60 | 62.40 | 45.3469 | 50.10 | 44.30 | 8.38562 |
| **DSAP - Digital skills among active population** | 32 | 24.10 | 67.70 | 48.2562 | 48.30 | 48.05 | 8.80967 |
| Countries | Angola, Bangladesh, Bolivia, Cambodia, Cameroon, Cape Verde, Côte d'Ivoire, Egypt, El Salvador, Eswatini, Ghana, Honduras, India, Indonesia, Kenya, Kyrgyz Republic, Lao PDR, Mauritania, Moldova, Mongolia, Morocco, Nicaragua, Nigeria, North Macedonia, Pakistan, Philippines, Senegal, Tunisia, Ukraine, Vietnam, Zambia, Zimbabwe ||||||| 

## 4.2. Methodology

This study investigates the impact of R&D on economic growth and how specific institutions influence this relationship. For this purpose, we use the structural equation modeling (SEM) method**.** SEM is a technique which depicts the relationship between different independent and a target (dependent) variable through a measurement model and a path diagram. A key advantage of SEM is its ability to test and estimate relationships between constructs. It also helps the assessment of construct validity by utilizing multiple measures for each construct and estimating measurement error, something other linear models do not (Weston & Gore, 2006). SEM uses confirmatory factor analysis (CFA) in order to measure the latent variables (factors) and item weights (factor loadings).

## 5. Results

After collecting data from the data source and identifying the model, the next step is to analyze the results obtained from the tests performed. This chapter discusses all the statistical methods that were performed in order to test the measurement and structural part of the analysis.

## 5.1. Confirmatory factor analysis (CFA)

As a preliminary test, we use CFA, which confirms where the latent variables lie and how much the variance between them is. Figure 2 shows the value of the correlation coefficients among the variables used for the factor analysis, as well as the aggregated variables. The Chi-Square Test value is 935.701 and the p-value is less than 0.001, so we can reject the null hypothesis and infer that there is a relationship between the variables. The KMO is 0.859, which is considered a good value. The Cronbach's Alpha values are above 0.7 (Table 3), which is also a good value (Samuels, 2015).

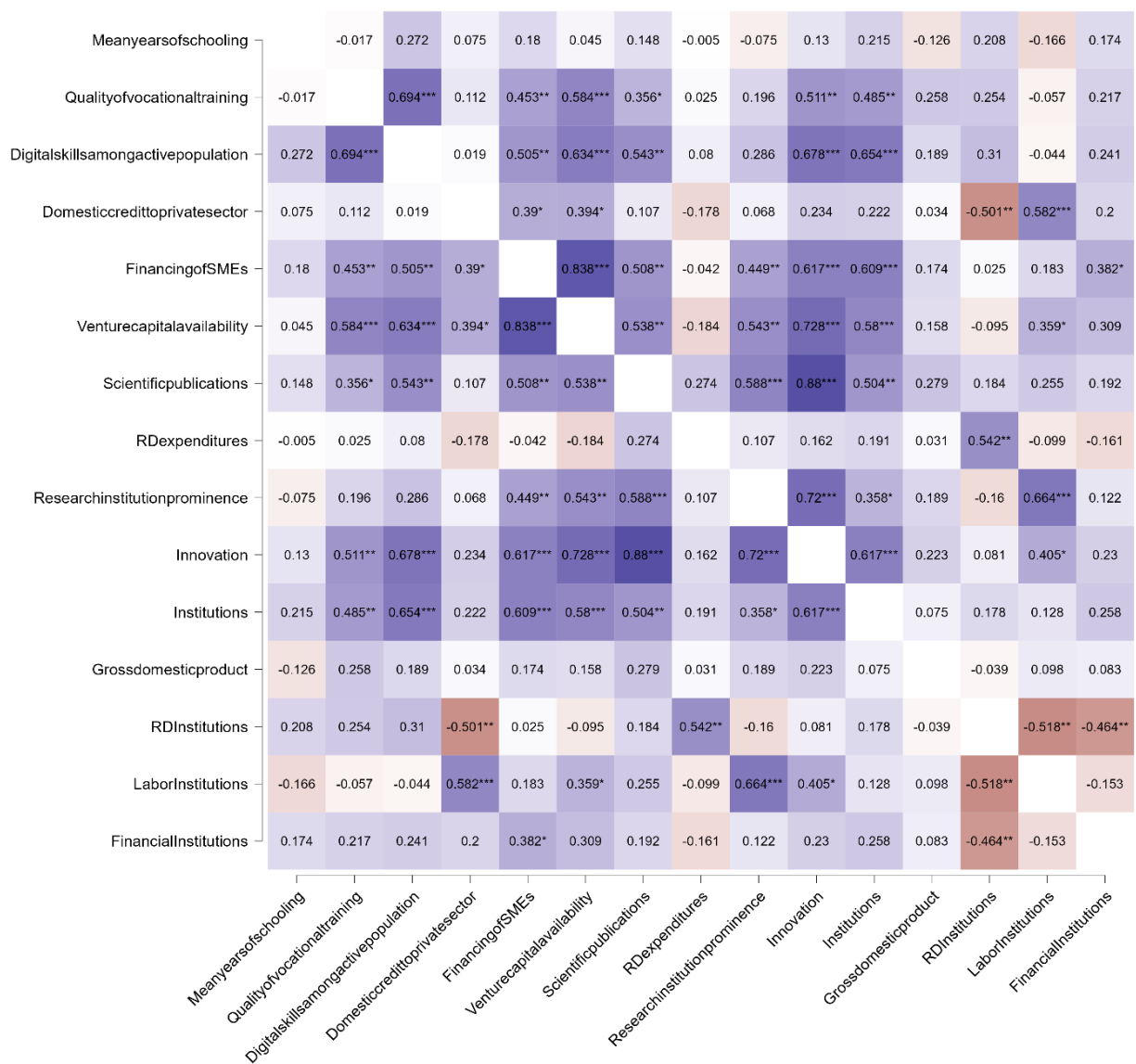

*Figure 2: The correlation map of the investigated variables*
*(Source: own work, generated with JASP)*

Table 3: Reliability analysis of the confirmatory factor analysis
*(Source: own calculations)*

| Factor | Coefficient ω | Coefficient α |
|---|---|---|
| RDI – R&D construct | 0.789 | 0.752 |
| FI – Capital Market construct | 0.789 | 0.771 |
| LI – Labor Market construct | 0.771 | 0.765 |
| Total | 0.845 | 0.879 |

After confirming the relationship between the observed and the latent variables through CFA, a path diagram is required to prove the structural relationships and assess model fit, confirming their effects on the possible outcome.

## 5.2. Path diagram

The path diagram of SPSS Amos Graphics offers a number of noteworthy features. To begin with, it includes both observed (manifest) and unobserved (latent) variables. Second, it includes both causal linkages between unobserved variables (represented by single-headed arrows) and bi-directional or correlational relationships between many residuals (e1, e2,….e15). The residual correlations account for the additional shared variance.

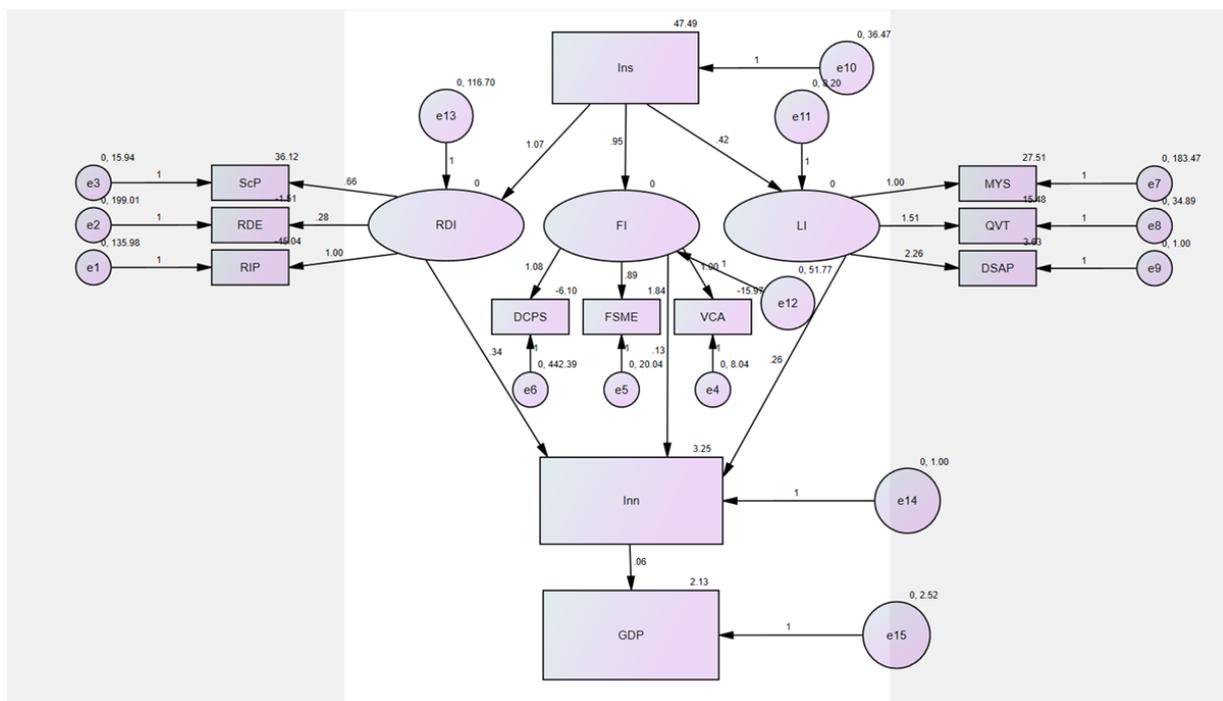

**Legend**

**Ins**: Institutions (see Table 1)
**RDI**: Research & Development construct built from the three indicators below (see Component variables section)
**ScP**: Scientific publications
**RDE**: R&D expenditure
**RIP**: Research institution prominence
**FI**: Capital market construct built from the three indicators below (see Component variables section)
**DCPS**: Domestic credit to private sector
**FSME**: Financing of SMEs

*VCA*: Venture capital availability
*LI*: Human capital construct built from the three indicators below (see Component variables section)
*MYS*: Mean years of schooling
*QVT*: Quality of vocational training
*DSAP*: Digital skills among active population
*Inn*: Innovation (see Table 1)
*GDP*: Average annual GDP growth 2011-20 (see Table 1)

Figure 3: SEM Path Diagram
(*Source: own work*)

Our theoretical framework assumes significant relationships between specific variables and innovation, subsequently influencing economic growth. Hypothesis 1 proposes a positive association between R&D performance (RDI) and innovation. Hypotheses 2 and 3 suggest that Capital market institutions (FI) and Labor market institutions (LI), respectively, positively impact innovation. Finally, Hypothesis 4 hypothesizes a positive correlation between innovation (Inn) and economic growth (GDP).

To empirically test these hypotheses, we employed a SEM approach, visualized in the path diagram (Figure 3). Latent variables representing RDI, FI, and LI were constructed using factor analysis. The SEM model assesses the direct and indirect effects of these latent variables on innovation, and subsequently, on economic growth. Significant positive relationships among these constructs would provide empirical support for our hypothesized model.

Table 4: Regression weights
(*Source: Author's own calculations*)
*Regression weight represents the expected change in dependent variable because of an increase in independent variable of one of its standardized units with all other independent variables unchanged* (Siegel & Wagner, 2022).

| Regression pairs | | | Estimate | S.E. | C.R. | P |
|---|---|---|---|---|---|---|
| RDI (R&D) | <--- | Institutions | 1.074 | .384 | 2.795 | .005 |
| FI (CM – Capital Market) | <--- | Institutions | .949 | .229 | 4.146 | *** |
| LI (HC – Human Capital) | <--- | Institutions | .415 | .281 | 1.480 | .139 |
| Innovation | <--- | RDI | .344 | .062 | 5.535 | *** |
| Innovation | <--- | FI | .128 | .047 | 2.698 | .007 |
| Innovation | <--- | LI | .262 | .200 | 1.312 | .190 |
| Research institution prominence | <--- | RDI | 1.000 | | | |
| RD expenditure | <--- | RDI | .282 | .211 | 1.338 | .181 |
| Scientific publications | <--- | RDI | .663 | .127 | 5.210 | *** |
| Venture capital availability | <--- | FI | 1.000 | | | |
| Financing of SMEs | <--- | FI | .890 | .134 | 6.622 | *** |
| Domestic credit to private sector | <--- | FI | 1.076 | .437 | 2.464 | .014 |
| Mean years of schooling | <--- | LI | 1.000 | | | |
| Quality of vocational training | <--- | LI | 1.515 | 1.014 | 1.494 | .135 |
| Digital skills among active population | <--- | LI | 2.263 | 1.457 | 1.553 | .120 |
| GDP growth | <--- | Innovation | .063 | .053 | 1.192 | .233 |

Table 3 shows the unstandardized coefficients and related test statistics. For every standardized unit change in the predictor, the degree of change in the dependent or mediating variable is represented by the unstandardized regression coefficient. Under the P column is the probability value corresponding to the null hypothesis that the test is zero. A p-value that is higher than 0.05 suggests that there may not be a significant relationship between the variables included in the model (Dahiru, 2011). The level of institutional development measured by the WEF has a significant impact on R&D and FI. A change of one standardized unit in the Institutions indicator increases both variables by about one unit (1.074 in case of R&D, and 0.949 in case of FI).

The regression weight of 0.344 indicates a positive relationship between RDI and innovation. For every standardized one-unit increase in RDI, innovation is expected to increase by approximately 0.344 units, assuming all other factors remain constant. The critical ratio of 5.535 suggests that the relationship between the innovation and RDI is statistically significant. Thus, we can reject the null hypothesis that the regression coefficient is equal to zero, indicating that R&D does have a significant effect on innovation. Middle-income countries with higher R&D expenditures and more scientific publications are likely to have higher levels of innovation. There is a strong, positive correlation between R&D and research institution prominence, indicating a strong association between R&D activities and prominent research institutions.

We have also detected a positive relationship between FI and innovation. For every standardized one-unit increase in FI, innovation is expected to increase by approximately 0.128 units, assuming all other factors remain constant. The relationship is significant at all standard levels (p=0.007). Venture capital availability, the financing of SMEs, and domestic credit availability are all significantly associated with FI and therefore have an influence on innovation, according to our model calculations.

The model points to a positive connection between LI and Innovation (0.262), as well as between innovation and economic growth (0.063), but these connections are not statistically significant (p=0.19 & 0.233).

### 5.2.1. Model fit summary

The results of the model fit tests can be found in Annex 1. The SEM model displays a good fit to the data. Key fit indices, such as the Chi-Square Test (CMIN/DF, 1.215 for default model), fit indices (NFI=0.756, RFI=0.690, IFI=0.946, TLI=0.926, CFI=0.942), and parsimony measures (PRatio=0.788, PNFI=0.595, PCFI=0.742) indicate acceptable to excellent model fit. Additionally, the root-mean-square error of approximation (RMSEA=0.08) and the index of model fit (FMIN=2) values were within acceptable ranges. The expected cross-validation index (ECVI=4.48) values indicate potential limitations in predicting future data and the Hoelter test suggests that the sample size is sufficient to support the model's significance.

SEM analysis typically requires large samples. Westland (2010) extensively studied sample size requirements, often determined by the ratio of indicators to latent variables. Our ratio of 4 suggests a sample size of 100. However, normally distributed variables allow for smaller samples. With the exception of "Research institution prominence" and "R&D expenditure," our variables are normally distributed (Shapiro-Wilk test). This normality strengthens the reliability of our model despite the smaller sample size.

## 6. Discussion

SEM analysis was used to test the theoretical framework, and the results indicate a good model fit. The SEM path diagram shows a significant positive relationship between institutions and R&D, as well as between institutions and capital markets (the relationship with human capital is also positive, but not significant; p=0.139). To examine the impact of different types of institutions on economic growth, we adopted an ecosystem approach in which different institutional factors, including those related to R&D, human capital and capital markets, collectively impact the innovation in a country, which in term affects economic growth. We found significant positive relationships between R&D and innovation and also between capital markets and innovation. However, the relationships between human capital and innovation and between innovation and economic growth were not significant. Thus, based on our research we find support for Hypotheses 1 and 2, while Hypotheses 3 and 4 were not confirmed.

The results support the first hypothesis, which claimed that there is a positive and significant relationship between R&D and innovation. Specifically, the estimate results show that a one-unit increase in R&D is associated with a 0.34-unit rise in innovation. Therefore, the results support H1. This backs up the earlier research by Coe et al. (1997), Lichtenberg (2001) and Aghion et al. (2011) that emphasized the critical role of R&D investment in driving innovation.

Capital markets are found to have a significant and positive impact on innovation, as proposed in the second hypothesis and confirmed by the results. The estimates show that a one-unit increase in financial institutions is associated with a 0.12-unit rise in innovation. The p-value for this relationship is 0.007 which means that there is 0.7% probability that the financial institutions' impact on innovation is by chance. Therefore, H2 is supported by the model results. Transparency in the financial institutions increases investors' confidence and attracts future capital investment. The results are in line with evidence from early empirical research by Shaw (1973) and Degong et al. (2021) showing a positive correlation between innovation and financial liberalization. Capital markets are essential for fostering entrepreneurship by providing financial support to startups and innovative ventures (Lerner, 1999). Entrepreneurship, in its essence, drives innovation, as entrepreneurs are frequently pioneers in creating new products, services, and business approaches (Schumpeter, 2017). Furthermore, innovative firms often attract substantial investments from capital markets due to their potential for growth and profitability (Gompers et al., 2005). Thus, the interplay between capital markets, entrepreneurship, and innovation creates a dynamic framework that promotes economic expansion and technological progress.

The results do not support the third hypothesis: H3 could not be confirmed by our model. Our results indicate that the endogenous theories put forward by Lucas (1988) or Azariadis and Drazen (1990) may not be applicable to developing countries. While human capital is often perceived as a key driver of innovation, the correlation between the two is complex. Some research posits that increased human capital, characterized by higher levels of education and skills, can foster innovation by enhancing individuals' ability to devise and execute new ideas (Acemoglu, 2002). Conversely, other studies suggest that an excessive focus on human capital can stifle innovation by promoting conformity and discouraging risk-taking and creative thinking (Naylor & Florida, 2003). Jones and Williams (1998) found that while human capital is crucial for productivity, it does not always translate into enhanced innovation. Likewise, Bessen and Hunt (2007) observed that boosting human capital through educational investments did not consistently correlate with elevated innovation levels across various industries. Hence, while human capital undeniably contributes to innovation, its influence can be modulated by factors such as organizational culture and incentives, creating an intricate and multifaceted relationship between the two.

Our analysis failed to identify a statistically significant positive correlation between innovation performance and economic growth among lower-middle-income developing countries. This finding contributes to the growing body of literature on the middle-income trap (Kharas & Gill, 2020). Hannum (2024) provides empirical evidence for this phenomenon in two distinct groups of countries, one of which aligns with our research focus. This finding is consistent with the World Bank's 3i model of development outlined in the World Bank's latest report on the middle-income trap (World Bank, 2024). According to the 3i model, initial growth is driven by investment (the first 'i'). However, a growth trap may ensue if the infusion of new technologies (the second 'i') is not realized. Only after overcoming this initial trap does innovation (the third 'i') become a crucial driver of development.

Our results suggest that the relationship between innovation, its underlying factors, and long-term economic growth in developing countries is more complex than the NGT model implies. Lower-middle-income countries may prematurely prioritize R&D institutions (RDI) when, in fact, enhancing infusion capabilities should take priority. This could explain why R&D performance correlates with innovation activity without significantly impacting growth. Human capital development is found to be a more critical predictor of overcoming the trap in this group of countries (Cm et al., 2024).

Institutional factors also play a significant role. Middle-income countries face institutional and political challenges, particularly in upgrading productivity through human capital and innovation, which necessitate substantial investments in institutional capacity (Doner & Schneider, 2016; Mickiewicz, 2023; Song et al., 2023). Our findings may reflect a broader pattern of middle-income growth failure among developing countries.

## 7. Conclusion

Weak institutions hinder investment in capital markets and education, negatively impacting firm performance. This results in underinvestment in human capital, limiting skilled labor, R&D, and innovation. Consequently, poor firm performance depresses macroeconomic growth, perpetuating a negative cycle. This dynamic is explained by our study's R&D ecosystem framework.

The research was conducted on the lower-middle-income group of countries as categorized by the World Economic Forum for the year 2019. Confirmatory factor analysis was conducted as a preliminary step of SEM, to test the relationship between the dependent and the independent variables. The Chi-Square value is 1.215, which is the indication of a good fit.

Our research confirmed that the capital market institutions and research and development features have a positive impact on innovation. We found two significant relationships: R&D features (regression weight estimate: +0.34, p=0.001) and capital market institutions (regression weight estimate: +0.12, p=0.007). Middle-income countries with higher R&D expenditure and more scientific publications are likely to have higher levels of innovation. There is a strong, positive correlation between R&D and research institution prominence, indicating a strong association between R&D activities and prominent research institutions. Venture capital availability, the financing of SMEs, and domestic credit availability are all significantly associated with CM and therefore have an influence on innovation, according to our model calculations.

An economy must prioritize creating a strong financial system, strengthening the legal system, and having an honest and upright government. Deficiencies in institutional quality must be addressed by policymakers, especially concerning the burden of labor regulations. The literature has recognized the disparity in R&D capacity between developed and developing nations and has proposed formal models of technological diffusion. However, the disparity does not stop with R&D capacity. Rather, it is widespread and affects a great deal of other facets of the economic system. A prime example is the labor markets of many developing nations, which differ significantly from those of developed countries due to factors including structural rigidities, under-employment, and concealed unemployment.

Our model indicates that although there is a positive relationship between innovation and economic growth, it is not statistically significant. This result could be interpreted as a sign of a middle-income trap (a growth failure identified by some studies in developing countries) among the lower middle-income group of countries.

A key limitation of this study is the relatively small number of countries included (32). While SEM benefits from larger samples for enhanced reliability, our sample size was determined by the research aim: evaluating R&D ecosystems specifically within lower-middle-income countries. This smaller sample increases the likelihood of Type II errors (false negatives), meaning we may fail to detect existing relationships. Specifically, the probability of correctly rejecting a false null hypothesis (i.e., finding a true effect) at a 5% significance level is reduced with smaller samples. Although the normal distribution of our variables partially mitigates this risk, the potential for missing true effects remains. Exact fit tests such the Chi-Square Test are more accurate in determining model fit and the CMIN/DF (Chi-Square divided by degrees of freedom) value of our model (1.215) also suggests a good fit.

Our use of 2019 data may also be debatable. This decision was made so that we could focus on a period that was characterized by robust global growth and could also avoid the distortionary impacts of the COVID19 pandemic on our results.

Annex 1

**SEM analysis, model fit summary**
**(Source: own calculations)**

| Measure | Estimate | Threshold | Interpretation |
| --- | --- | --- | --- |
| CMIN | 63.165 | - | - |
| DF | 52 | - | - |
| CMIN/DF | 1.215 | 2 | Excellent |
| NFI | 0.756 | Closer to 1 | Good |
| RFI | 0.690 | 1= Perfect Fit<br><br>Closer to 1= Good fit | Good |
| IFI | 0.946 | 1= Perfect Fit<br><br>Closer to 1= Good fit | Good |
| TLI | 0.926 | 1= Perfect Fit<br><br>Closer to 1= Good fit | Good |
| CFI | 0.942 | 1 = Perfect fit<br>≥ 0.95 = Excellent fit<br>≥ .90 = Acceptable fit | Good |
| PRatio | 0.788 | No cutoff value | - |
| PNFI | 0.595 | 0.5 and above | Excellent |
| PCFI | 0.742 | 0.6 and above | Excellent |
| NCP | 0 to 35.39 | - | - |
| FMIN | 2 | No cutoff value | - |
| RMSEA | 0.08 | 0.05 or 0.08 | Excellent |
| AIC | 139.16 | - | - |
| BCC | 194.05 | - | - |
| ECVI | 4.48 | No cutoff value | - |
| MCVI | 6.26 | No cutoff value | - |
| Hoelter 0.5 | 35 | - | - |
| Hoelter 0.1 | 39 | - | - |